\documentclass[conference]{IEEEtran}
\IEEEoverridecommandlockouts
\usepackage[nolist]{acronym}
\usepackage{cite}
\usepackage{amsmath,amssymb,amsfonts}
\usepackage{algorithmic}
\usepackage{graphicx}
\usepackage{textcomp}
\usepackage{xcolor}
\usepackage[hyphens]{url}
\def\BibTeX{{\rm B\kern-.05em{\sc i\kern-.025em b}\kern-.08em
    T\kern-.1667em\lower.7ex\hbox{E}\kern-.125emX}}
\begin{document}

\title{Architecture Considerations for ISAC in 6G\\
\thanks{This work has received support from the Smart Networks and Services Joint Undertaking (SNS JU) under the European Union's Horizon Europe research and innovation programme under Grant Agreement No 101192521 (MultiX).}
}

\author{\IEEEauthorblockN{Sebastian Robitzsch\IEEEauthorrefmark{1}, Laksh Bhatia\IEEEauthorrefmark{1}, Konstantinos G. Filis\IEEEauthorrefmark{2}, Neda Petreska\IEEEauthorrefmark{3}, Michael Bahr\IEEEauthorrefmark{3},\\Pablo Picazo Martinez\IEEEauthorrefmark{4}, Xi Li\IEEEauthorrefmark{5}}
\IEEEauthorblockA{\IEEEauthorrefmark{1}InterDigital Europe Ltd, London, United Kingdom, Email: \{firstname.surname\}@interdigital.com}
\IEEEauthorblockA{\IEEEauthorrefmark{2}Hellenic Telecommunications Organization, Greece, Email: kfilis@ote.gr}
\IEEEauthorblockA{\IEEEauthorrefmark{3}Siemens AG, Munich, Germany, Email: \{neda.petreska, bahr\}@siemens.com}
\IEEEauthorblockA{\IEEEauthorrefmark{4}Universidad Carlos III de Madrid, Leganes, Spain, Email: papicazo@pa.uc3m.es}
\IEEEauthorblockA{\IEEEauthorrefmark{5}NEC Laboratories Europe, Germany, Email: xi.li@neclab.eu}}

\maketitle

\begin{abstract}
ISAC is emerging as a foundational capability in 6G, enabling mobile networks to not only offer communication services but also to sense and perceive their environment at scale. This paper explores architectural considerations to enable sensing in 6G, extending on recent developments by (pre-)standardisation bodies such as 3GPP and ETSI. Selected ISAC use cases are presented from the European MultiX project including associated potential functional system requirements. The paper proposes a 6G system architecture that integrates newly proposed NFs for the purpose of sensing and demonstrates how they are being used in offering sensing as a service. Protocol stack adaptations for both control and a newly proposed sensing plane are discussed.
\end{abstract}

\begin{IEEEkeywords}
\acl{ISAC}, \acl{6GS}, \acl{SA}, 3GPP, ETSI, Standardisation.
\end{IEEEkeywords}

\acrodef{5G-ACIA}{5G Alliance for Connected Industries and Automation}
\acrodef{6GS}{6G System}

\acrodef{AAA}{Authentication, Authorisation and Accounting}
\acrodef{AF}{Application Function}
\acrodef{AI}{Artificial Intelligence}
\acrodef{AMF}{Access and Mobility Function}
\acrodef{AN}{Access Network}
\acrodef{AN-SC}{\ac{AN}-Sensing Control}
\acrodef{AP}{Access Point}
\acrodef{API}{Application Programming Interface}

\acrodef{BS}{Base Station}

\acrodef{CN}{Core Network}
\acrodef{CP}{Control Plane}
\acrodef{CPE}{Customer Premises Equipment}
\acrodef{CSI}{Channel State Information}
\acrodef{CU}{Central Unit}

\acrodef{DN}{Data Network}
\acrodef{DT}{Digital Twin}
\acrodef{DU}{Distributed Unit}

\acrodef{EC}{European Commission}
\acrodef{eMBB}{Enhanced Mobile Broadband}
\acrodef{EPC}{Evolved Packet Core}
\acrodef{ETSI}{European Telecommunication Standardisation Institute}

\acrodef{GDPR}{General Data Protection Regulations}
\acrodef{gNB}{gNodeB}
\acrodef{GPRS}{General Packet Radio Service}
\acrodef{GR}{Group Report}
\acrodef{GTP-U}{\ac{GPRS} Tunnelling Protocol - User Plane}

\acrodef{HRLLC}{Hyper Reliable Low Latency Communication}
\acrodef{HTTP/3}{Hypertext Transfer Protocol Version 3}

\acrodef{IMT}{International Mobile Telecommunications}
\acrodef{IP}{Internet Protocol}
\acrodef{ISAC}{Integrated Sensing and Communication}
\acrodef{ISG}{Industry Specification Group}
\acrodef{ITU-R}{International Telecommunication Union - Radiocommunication Sector}
\acrodef{I/Q}{In-phase and Quadrature}

\acrodef{KPI}{Key Performance Indicator}

\acrodef{LiDAR}{Light Detection and Ranging}
\acrodef{LOS}{Line of Sight}

\acrodef{MASQUE}{Multiplexed Application Substrate over QUIC Encryption}
\acrodef{MM}{Mobility Management}
\acrodef{MNO}{Mobile Network Operator}
\acrodef{mMTC}{Massive Machine Type Communication}

\acrodef{N3IWF}{Non-3GPP Interworking Function}
\acrodef{NAS}{Non-Access Stratum}
\acrodef{NGAP}{Next-Generation Application Protocol}
\acrodef{NF}{Network Function}
\acrodef{NSA}{Non-Standalone}

\acrodef{PCF}{Policy Coordination Function}
\acrodef{PFR}{Potential Functional Requirement}

\acrodef{QoS}{Quality of Service}

\acrodef{RAN}{Radio Access Network}
\acrodef{RIS}[RIS]{reconfigurable intelligent surface}

\acrodef{SA}{System Architecture}
\acrodef{SA2}{\ac{SA} Working Group 2}
\acrodef{SBA}{Service-Based Architecture}
\acrodef{SBI}{Service-Based Interface}
\acrodef{SC}{Sensing Coordination}
\acrodef{SCF}{Sensing Coordination Function}
\acrodef{SCTP}{Stream Control Transmission Protocol}
\acrodef{SD}{Sensing Data}
\acrodef{SDP}{Sensing Data Producer}
\acrodef{SEF}{Sensing Exposure Function}
\acrodef{SEF-SC}{\ac{SEF}-Sensing Control}
\acrodef{SM}{Session Management}
\acrodef{SMF}{Session Management Function}
\acrodef{SP}{Sensing Plane}
\acrodef{SPF}{Sensing Processing Functions}
\acrodef{SR}{Sensing Result}
\acrodef{SRP}{Sensing Result Producer}
\acrodef{SPF-SC}{\ac{SPF}-Sensing Control}
\acrodef{SRv6}{Segment Routing Version 6}
\acrodef{SSC}{Sensing Service Consumer}
\acrodef{SRX}{Sensing Receiver}
\acrodef{ST}{Sensing Task}
\acrodef{STG}{\ac{ST} Group}
\acrodef{STID}{\ac{ST} Identifier}
\acrodef{STX}{Sensing Transmitter}

\acrodef{TCP}{Transmission Control Protocol}
\acrodef{TNAN}{Trusted Non-3GPP Access Network}
\acrodef{TR}{Technical Report}
\acrodef{TS}{Technical Specification}
\acrodef{TSSA}{Target Sensing Service Area}

\acrodef{UDM}{Unified Data Management}
\acrodef{UE}{User Equipment}
\acrodef{UP}{User Plane}
\acrodef{UDP}{User Datagram Protocol}
\acrodef{UPF}{\ac{UP} Function}
\acrodef{URLLC}{Ultra Reliable Low Latency Communication}

\acrodef{WG}{Working Group}

\section{Introduction}
Back in November 2023, \ac{ITU-R} released their \ac{IMT} 2030 vision framework \cite{imt2030} which provides six usage scenarios and four overarching aspects for the usage scenarios to build upon. Three usage scenarios out of the six were taken from \ac{ITU-R}'s 2020 vision, i.e. immersive communication (adopted from \ac{eMBB} in 5G), massive communication (adopted from \ac{mMTC} in 5G) and \ac{HRLLC} (adopted from \ac{URLLC} in 5G). Additionally, \ac{ITU-R} spell out three new usage scenarios in their 2030 vision, i.e. ubiquitous connectivity, \acs{AI} and Communication, and \ac{ISAC}. This paper focuses on the \ac{ISAC} usage scenarios and provides \ac{SA} considerations to enable this usage scenario in \acp{6GS} covering architectural and protocol \ac{SA} aspects. 

3GPP \ac{TR} 22.837~\cite{3gpp22837} provides consolidated potential functional requirements for \ac{ISAC} categorised into general, configuration and authorisation, network exposure, security and charging. The performance requirements are provided as numerical values, clustered around three scenarios, specifically object detection and tracking, environment monitoring, and motion monitoring. Sensing specific performance requirements include confidence level, accuracy of positioning, accuracy of velocity, sensing resolution, maximum sensing service latency, refresh rate, missed detection and false alarm. 3GPP then specifies the service requirements for \ac{ISAC}-enabled 5G-Advanced systems in \ac{TS} 22.137~\cite{3gpp22137}, which provides terminology definitions for sensing and defined sensing operations in 3GPP systems supporting co-located and separated sensing receiver and  transmitter and the processing of 3GPP sensing data inside the 3GPP system for objects outside of the 3GPP system. 

The remainder of the paper is structured as follows: Section~\ref{sec:usecases} provides a brief description and \acp{PFR} of \ac{ISAC} use cases considered in the MultiX project \cite{multix} and identified as not covered in \ac{ETSI} \cite{etsi-isg-gr001} or 3GPP~\cite{3gpp22137, 3gpp22837}. This is followed by architectural considerations to enable \ac{ISAC} in \aclp{6GS} in Section~\ref{sec:archconsiderations}, providing foundational assumptions, an overview of what a \acl{ST} is composed of in \acp{6GS} and a revised \ac{SA}. The paper is concluded in Section~\ref{sec:conclusions}.

\section{Considered Use Cases}\label{sec:usecases}
In 2022, 3GPP \ac{SA}1 started working on \ac{TR} 22.837 as part of Release 19 which studied service requirements for \ac{ISAC} in 5G in the form of use cases~\cite{3gpp22837}. The technical report was finalised in mid-2024 and provides in total 32 \ac{ISAC} use cases covering a wide range of applications, including object and intruder detection, collision avoidance and trajectory tracking, automotive navigation, public safety, rainfall monitoring, health and sports monitoring. Special considerations are provided on aspects of confidentiality, integrity and privacy, as well as regulatory requirements. This can be explained when comparing \ac{ISAC} with technical advances in previous evolution of mobile networks, i.e., \ac{ISAC}-enabled networks offer the ability to track and identify objects in the environment including their micro movements such as heart rates. If the resolution of sensing data is sufficiently high, sensing results could theoretically include visual profiles of humans and critical infrastructures; and as it is foreseen that sensing results may be exposed to trusted or untrusted third parties, causing data privacy and security concerns. Thus, 3GPP \ac{TR} 22.837 spells out potential requirements around encryption, integrity protection, \acs{GDPR} adherence and authorisation when developing technical specifications that enable \ac{ISAC} in mobile networks. 

In late 2024, 3GPP \ac{SA}1 then started the study on 6G use cases and service requirements as a Release 20 technical report, i.e., \ac{TR} 22.870~\cite{3gpp22870}. This report provides considerations for system and operational aspects around migration to 6G, including security, sustainability and energy efficiency, and device support. These use cases are clustered around the \ac{IMT} 2030 usage scenarios, i.e., \ac{AI}, \ac{ISAC}, Ubiquitous Connectivity and Massive Communication. This report also provides further use cases on industry and verticals. In the area of \ac{ISAC}, the report comprises 13 approved use cases at the time of writing this article. The key differences between 5G-Advanced and 6G \ac{ISAC} use cases in 3GPP \ac{SA}1 are more stringent performance requirements and functional requirements beyond what was provided in \ac{TR} 22.837~\cite{3gpp22837}. 

In 2024, the \ac{ETSI} \ac{ISG} on \ac{ISAC} released a \ac{GR} 6G \ac{ISAC} use cases~\cite{etsi-isg-gr001}. The \ac{GR} describes 18 advanced use cases ranging from human motion recognition and emergency rescue to autonomous vehicle navigation and industrial robotics. The \ac{GR} also provides functional requirements and proposes new \acp{KPI} such as fine motion accuracy and sensing service range, ensuring a robust performance evaluation framework for 6G sensing services. Some of the 3GPP 6G use cases in \ac{TR} 22.870 on \ac{ISAC} are presented in a more elaborate format in the \ac{ISG}'s \ac{GR}. Also the \acs{5G-ACIA} collected \ac{ISAC} use cases focussing on industrial applications \cite{5g-acia-isac}.

The following three \ac{ISAC} use cases presented in this paper have been developed as part of the MultiX \ac{EC}-co-funded research project \cite{multix} and form the basis to demonstrate the importance of \ac{ISAC} in traditional vertical domains on industrial manufacturing and entertainment as well as exploring new domains around value-added services through \ac{ISAC}. Furthermore, the use cases spell out functional system requirements providing a \ac{6GS} must provide in order to enable these use cases. Those functional system requirements are referenced back in the architectural considerations section.

\subsection{Smart Shopping Tracker}\label{sec:usecases-uc1}

\begin{figure}[!t]
\centering
\includegraphics[width=\linewidth]{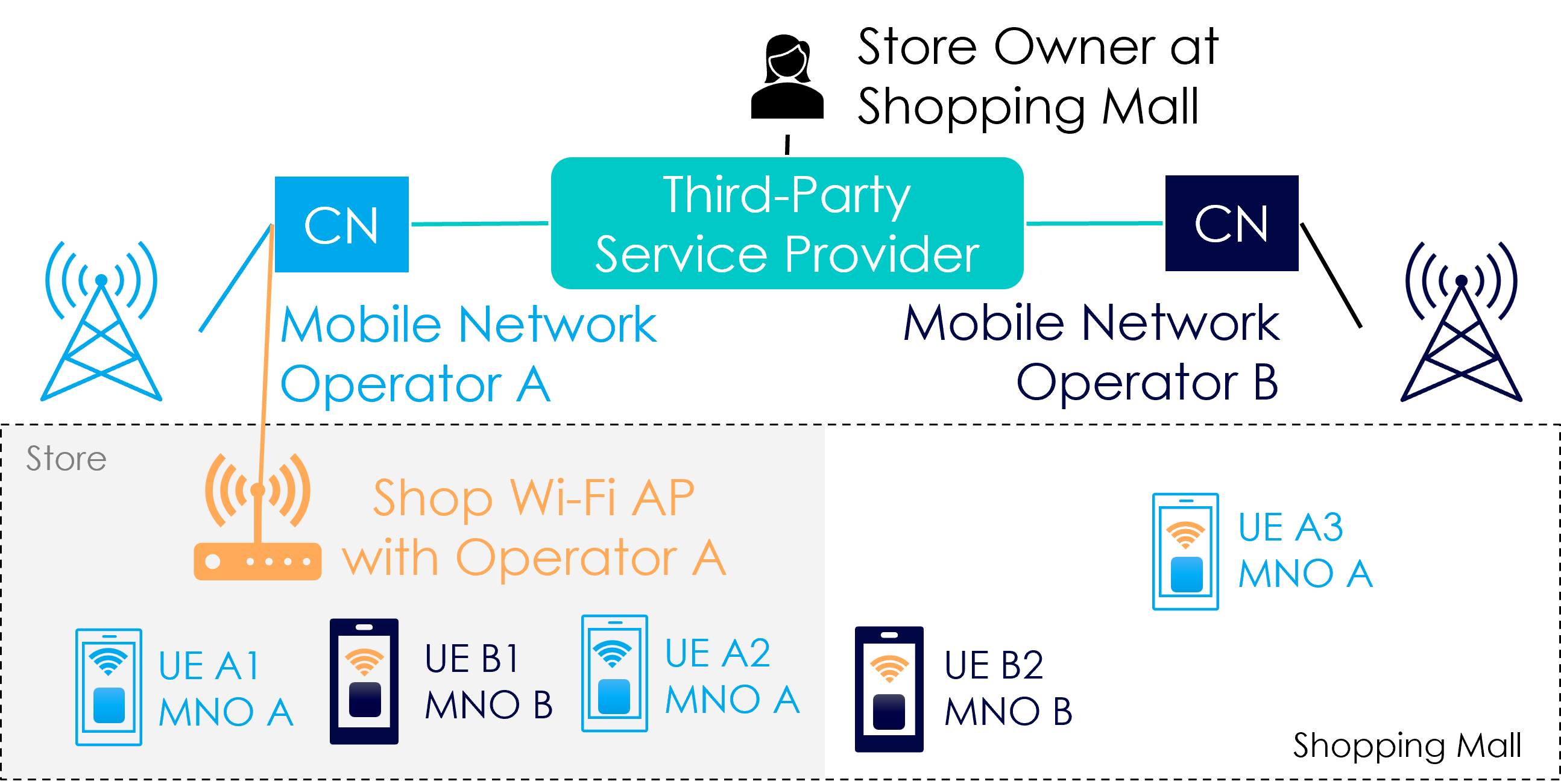}
\caption{Illustration of Smart Shopping Tracker Scenario Utilising Sensing Results from Multiple \aclp{MNO}.}
\label{fig:uc-1}
\end{figure}
This use case describes the ability for a third-party service provider to collect sensing results from multiple \acp{MNO} leveraging 6G and non-6G sensing data. The Smart Shopping Tracker demonstrates the business opportunities of \ac{ISAC} for \acp{MNO} and third-party service providers to offer novel services to existing sectors, e.g., the retail sector. The proposition is that a shop owner at a larger shopping mall is interested in customers' shopping behaviour. The store owner finds out about a company called SENSOR - not affiliated with the shopping mall or with any particular \ac{MNO} - but offers a service to retailers to obtain customer shopping behaviour insights. This is conveyed as a 3D representation of the store's interior and a heatmap-like overlay showing the customer interest. The scenario is illustrated in \figurename~\ref{fig:uc-1}. It is assumed that SENSOR has contracts with both \acp{MNO} have coverage at the shopping mall and have \ac{ISAC} capabilities in their 6G networks. The retailer also acquired a broadband connection from \ac{MNO} A and has a sensing-enabled Wi-Fi \ac{AP} in their shop. A number of store visitors are with either \ac{MNO} A or B and their 6G devices are IEEE 802.11bf-enabled as well. 

\begin{table}[!hbt]
\renewcommand{\arraystretch}{1.3}
\caption{Identified \acp{PFR} for the Shopping Experience Tracker Use Case}
\label{tab:uc1-pfrs}
\centering
\begin{tabular}{p{.1\linewidth}||p{0.8\linewidth}}
\hline
\bfseries PFR & \bfseries Description\\
\hline\hline
PFR1-1& 6G Network should be able to collect 
sensing-related capabilities from \acp{UE} with 6G and non-6G sensing functionalities\\
PFR1-2& 6G Network should be able to 
configure 6G \acp{UE} with sensing-enabled Wi-Fi 
STA functionalities to perform sensing\\
PFR1-3& 6G Network should be able to decide 
whether 6G or non-6G sensing should be 
performed\\
PFR1-4& 6G Network should be able to switch 
between 6G and non-6G sensing while executing a sensing activity\\
PFR1-5& 6G Network should be able to fuse 6G 
and non-6G sensing data into sensing results\\
PFR1-6&6G Network should be able to expose 
sensing results to one or more third party 
service providers and be charged accordingly\\
\hline
\end{tabular}
\end{table}

The shop owner booked the service from SENSOR to obtain information about which products were of most interest to their customers. Once the shop owner had completed their contractual agreement, SENSOR issues a sensing service request to both \acp{MNO} providing the necessary information about the area to be sensed (e.g. cartesian coordinates of the shop outline), the desired \acp{KPI} for the sensing results, and the daily opening times of the store when sensing results should be provided. At store opening times, both \acp{MNO} start setting up their sensing activities to provide the sensing results SENSOR requested. Using the provided location of the store, both \acp{MNO} check for available \acp{UE} in the sensing area for 6G \ac{UE}-involved mono-, bi- and multi-static sensing. While continuously assessing the \acp{KPI} of the generated sensing results (mainly around target object identification confidence level and resolution accuracy), \ac{MNO} B's network determines that the sensing result \acp{KPI} have not been met and consults a trusted \ac{AF} for assistance information to utilise Wi-Fi sensing of \acp{UE}, provisioned by \ac{MNO} B's shops, located inside the sensing area. Using the identifiers of the \acp{UE} provided by the \ac{MNO}'s network, the \ac{AF} reaches out to each \ac{UE} that is capable of performing Wi-Fi sensing to provide results using the shopping mall's deployed Wi-Fi-enabled \acp{AP}. In contrast, \ac{MNO} A has the ability to complement 6G sensing data with Wi-Fi sensing data from the Wi-Fi \ac{AP} the store deployed to offer a potentially better sensing service continuity.
Table~\ref{tab:uc1-pfrs} lists and describes the identified \acp{PFR} for this use case.

\subsection{Smart Home}
This use case exemplifies how \ac{ISAC} technology can create new business opportunities for \acp{MNO} and service providers allowing them to offer new services for the smart home sector. The proposition is that a home owner wishes to be informed of intruders (person or harmful animal), outside or inside their private property. Currently, this is achieved by utilising IP/CCTV cameras, passive infrared sensors and microwave radars capable of monitoring and detecting objects and/or motion. However, these sensors either require \ac{LOS} or their monitoring area capabilities are limited and they can be easily accessed and damaged by malicious intruders. Wireless signals provided by 6G \acp{BS}, Wi-Fi \ac{CPE} and \acp{UE} with sensing capabilities make it possible to monitor wider areas in a more robust and secure manner. In addition, the fusion of the sensing data of 6G and non-6G wireless devices can improve accuracy and ubiquity of the intruder detection service.

In this use case, it is considered that a home owner is a subscriber to a certain telecom provider and home owner is informed about a new security service, called SENSE, which can provide them with a virtual security system that does not require the home owner to install any equipment or sensors to their premise. SENSE is based on the new sensing capabilities provided by the telecom provider (6G \acp{BS}, Wi-Fi \acp{CPE} and 6G \acp{UE}). With SENSE, the home owner will receive tamper-proof sensing information from the outdoor base stations, covering the entire outdoor area with high accuracy and no blind spots. The telecom provider will also upgrade the home owner’s Wi-Fi \ac{CPE} and \acp{UE} with ones supporting sensing functionality, allowing them to receive high-accuracy sensing information also for the indoor of the house, again with high accuracy and no blind spots. Therefore, the home owner will be able to be notified of both outdoor and indoor intruders knowing their exact location. 

\begin{figure}[!t]
\centering
\includegraphics[width=0.9\linewidth]{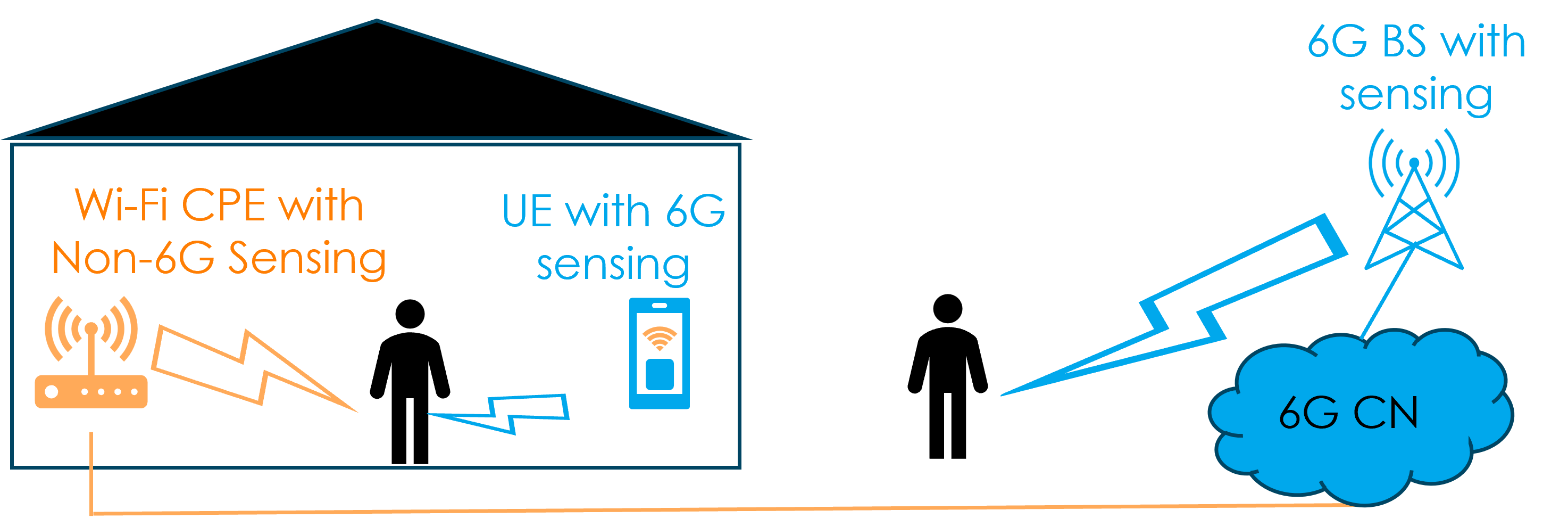}
\caption{Smart Home Use Case.}
\label{fig:6g-uc2}
\end{figure}

The scenario of how the new service SENSE works is illustrated in \figurename~\ref{fig:6g-uc2}. 
Outdoor coverage is provided by a 6G \ac{BS} with sensing capabilities while indoor coverage is provided by a Wi-Fi \ac{CPE} and one or more \acp{UE}. The SENSE service will also be able to support the home owner’s 6G and Wi-Fi wireless cameras (optional), both outside and inside the premises. The home owner will be able to designate the outdoor surveillance area, so whenever the sensing function of the BS detects an intruder within the surveillance area, the home owner will be notified via push notifications or SMS so that they can decide on their next actions. Indoor intruders can be detected by the sensing function of the Wi-Fi \ac{CPE} and/or \acp{UE}. Due to the motion of an indoor human, the 6G sensing signal measured by the \acp{UE} and the sensing signal of the Wi-Fi \ac{CPE} will be influenced. By analysing and collecting the sensing information, such as Doppler frequency shift, amplitude or phase change, the behaviour of an indoor human could be detected. 
Similar services can also be provided by third party service providers who can obtain the necessary sensing data, after signing contracts with one or more telecom operators. Table~\ref{tab:uc2-pfrs} lists and describes the identified \acp{PFR} for this use case.

\begin{table}[!hbt]
\renewcommand{\arraystretch}{1.3}
\caption{Identified \acp{PFR} for the Smart Home Use Case}
\label{tab:uc2-pfrs}
\centering
\begin{tabular}{p{.1\linewidth}||p{0.8\linewidth}}
\hline
\bfseries PFR & \bfseries Description\\
\hline\hline
PFR2-1& The 6G Network should be able to configure and control sensing-enabled \acp{BS}, Wi-Fi \acp{CPE} and \acp{UE}\\

PFR2-2& The 6G Network should provide a mechanism for an \ac{MNO} to authorise a \ac{UE} for sensing\\

PFR2-3& The 6G network should be able to collect 6G and non-6G sensing-related information from sensing-enabled \acp{BS}, Wi-Fi \acp{CPE} and \acp{UE}\\

PFR2-4& The 6G Network should be able to decide whether 6G or non-6G sensing should be performed\\

PFR2-5& The 6G Network should be able to fuse 6G and non-6G sensing data\\

PFR2-6& The 6G Network should be able to expose sensing results to one or more third party service providers\\
\hline
\end{tabular}
\end{table}

\subsection{Collaborating Robots in Smart Factories}
In this use case, robots work collaboratively on an industrial task. They are supported by means of a multi-layer \ac{DT} and of sensing capabilities of different kinds. Each robot is equipped with specific sensors tailored to perform certain functions. However, the challenge lies in the fact that these robots, when operating individually, lack the comprehensive capability and information required to perform the industrial task successfully. The primary goal is to enable a group of collaborating robots to perform a task that surpasses their individual capabilities and to allow each robot to perceive its environment beyond its local sensory limitations. 

\begin{figure}[!t]
\centering
\includegraphics[width=\linewidth]{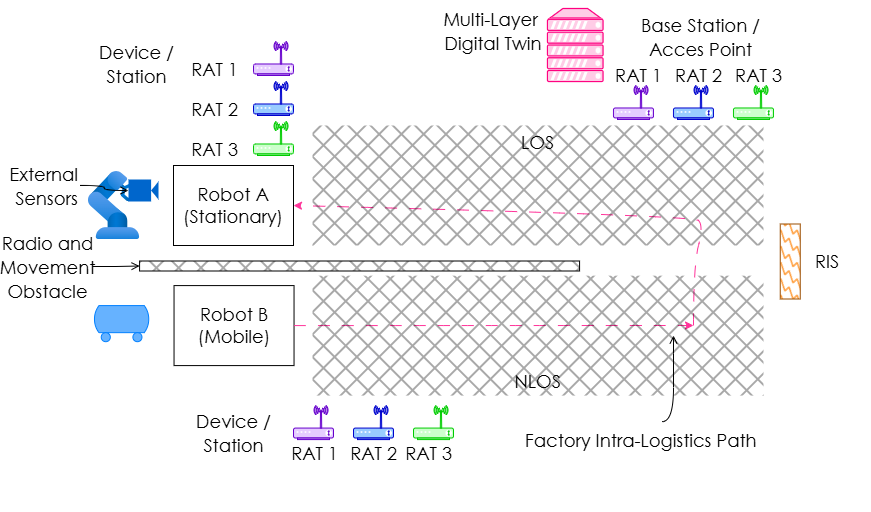}
\caption{Collaborating Robots in Smart Factories Use Case.}
\label{fig:6g-uc_siemens}
\end{figure}

The industrial environment delivers various sensor information, that provides context information for different areas (e.g. different robots may have different sensing ranges) and topics (e.g. sensors for ranging vs. sensors for measuring radio channel characteristics). The various sensor information may feed different \acp{DT}, such as an Environment \ac{DT}, a Network \ac{DT}, and an Application \ac{DT}. This is where the concept of a multi-layer \ac{DT} comes into play. Like the robots, each \ac{DT} on its own lacks the capability to provide the full context necessary for the industrial application or communication service. The multi-layer DT combines the individual \acp{DT} to create a comprehensive view, that enables the identification and resolution of potential issues, improves decision-making, and allows for dynamic adaptations and reconfiguration of the industrial application or network, based on specific circumstances and predicted events. Especially, it can provide the aggregated sensing results to the collaborating robots. This enhances the sensing capabilities of the individual robots and supports them in their collaboration and in other tasks that benefit from environmental information acquired by sensing methods.

\figurename~\ref{fig:6g-uc_siemens} shows an example scenario for collaborating robots in a smart factory. Robot B has to move to Robot A in order to work collaboratively on a task. The different local sensing results of both robots are aggregated in the multi-layer \ac{DT}. The aggregated sensing information provides guidance for the movement of Robot B, e.g. path routing, detection of obstacles, optimal connectivity. Robot B can now see beyond the wall that hides Robot A, it can determine a path around the wall, and the sensing information in the multi-layer \ac{DT} is used to provide reliable connectivity with the needed QoS by dynamically reconfiguring, for instance, the \acf{RIS} in \figurename~\ref{fig:6g-uc_siemens}.

The multi-layer \ac{DT} provides a comprehensive and dynamic virtual representation that empowers robots to collaborate by making joint decisions, adapts to changing conditions, and performs tasks beyond the robots' individual capabilities. This collaborative approach leads to improved efficiency, productivity, and reliability in industrial applications. Table~\ref{tab:uc3-pfrs} lists and describes major identified \acp{PFR} for this use case.

\begin{table}[!t]
\renewcommand{\arraystretch}{1.3}
\caption{\acp{PFR} for the Collaborative Robots in Smart Factories Use Case}
\label{tab:uc3-pfrs}
\centering
\begin{tabular}{p{.1\linewidth}||p{0.8\linewidth}}
\hline
\bfseries PFR & \bfseries Description\\
\hline\hline
PFR3-1& The (private) 6G network should be able to configure and control sensing-enabled network devices such as \acp{BS}, Wi-Fi \acp{CPE}, and \acp{UE}.\\
PFR3-2& The (private) 6G network should be able to collect 6G and non-6G sensing-related information from sensing-enabled network devices in a Multi-layer \ac{DT}.\\
PFR3-3& The (private) 6G network should be able to fuse 6G and non-6G sensing data.\\
PFR3-4& The (private) 6G network should be able to dynamically reconfigure network devices based on aggregated sensing information from the Multi-layer \ac{DT} in order to provide reliable connectivity with the required Quality of Service.\\
\hline
\end{tabular}
\end{table}
\section{Architectural Considerations}\label{sec:archconsiderations}
This section presents the architectural proposition on how to enable \ac{ISAC} in a \ac{6GS}. The section is structured around foundational assumptions for \ac{6GS}, followed by how sensing services are envisaged to be fulfilled by a \ac{6GS}. This baseline then is leveraged to provide architecture considerations on how to enable \ac{ISAC} in a \ac{6GS}, split between a refined and extended \ac{CP} and newly proposed \ac{SP}.

\subsection{Foundational Assumptions for \aclp{6GS}}\label{sec:archconsiderations-foundational}
While the architectural specification work has not started in 3GPP's \ac{SA2} \ac{WG}, a range of propositions have been made in \ac{EC}-funded research projects such as \cite{hexa-x-ii-d2.5, stylianopoulos2025distributedintelligentsensingcommunications} and pre-standardisation fora \cite{ngmn-6garch, snsju-archwg}. As a result of the cited work, several key architectural principles for \ac{6GS} can be identified as summarised hereafter:

\textbf{Modularity:} With 5G Rel.15, 3GPP adopted \ac{SBA} principles for its \ac{CN} which allows to deploy the majority of all 5G \acp{NF} in a truly cloud-native fashion. Such paradigm shift allows operators to scale a 5G \ac{CN} on-demand either based on load or failover scenarios of \ac{NF} instances. To achieve such scalable \ac{CN} 3GPP disintegrated 4G's \ac{EPC} from three \acp{NF} into \acp{NF} with a smaller set of functionality. Furthermore, such approach allows to decouple a technology enabler from other \acp{NF} and allows operators to not deploy such functionality, if not needed. Furthermore, such a modular approach allows functionality to be decoupled from each other and \acp{MNO} the flexibility to only deploy the functionality that interests them.  It is desired by operators to enable such scalable \ac{CN} for all \acp{NF} in 6G.

\textbf{Simplicity:} Operators have expressed their strong desire for 6G not to have multiple architectural options, as it was done for 5G with the two modes non-standalone and standalone. Building on top of a single 6G architecture, interface and protocol choices on all planes should be further unified paving the way for a fully cloud-native 6G system. 

\textbf{Sustainability:} The environmental impact should be further reduced in 6G, with a strong focus on energy consumption and architectural enablers to achieve that. 
Radio, networking, compute and storage resources should be able to utilise more efficiently and optimised on-demand where needed. Economic sustainability and environmental sustainability are the key priority for the architecture of future-proof 6G, which should be enabled by cost-effective deployment and operations with less energy consumed and efficient use of all types of resources. 

\textbf{Network-as-a-Service:} As in 5G, services offered by a \ac{6GS} should be exposed by a common \ac{API} with any newly added service offering around the \ac{IMT} 2030 usage scenarios leveraging the same principles.

\subsection{Sensing Service Workflow}\label{sec:archconsiderations-sensingworkflow}
\begin{figure}[!t]
\centering
\includegraphics[width=\linewidth]{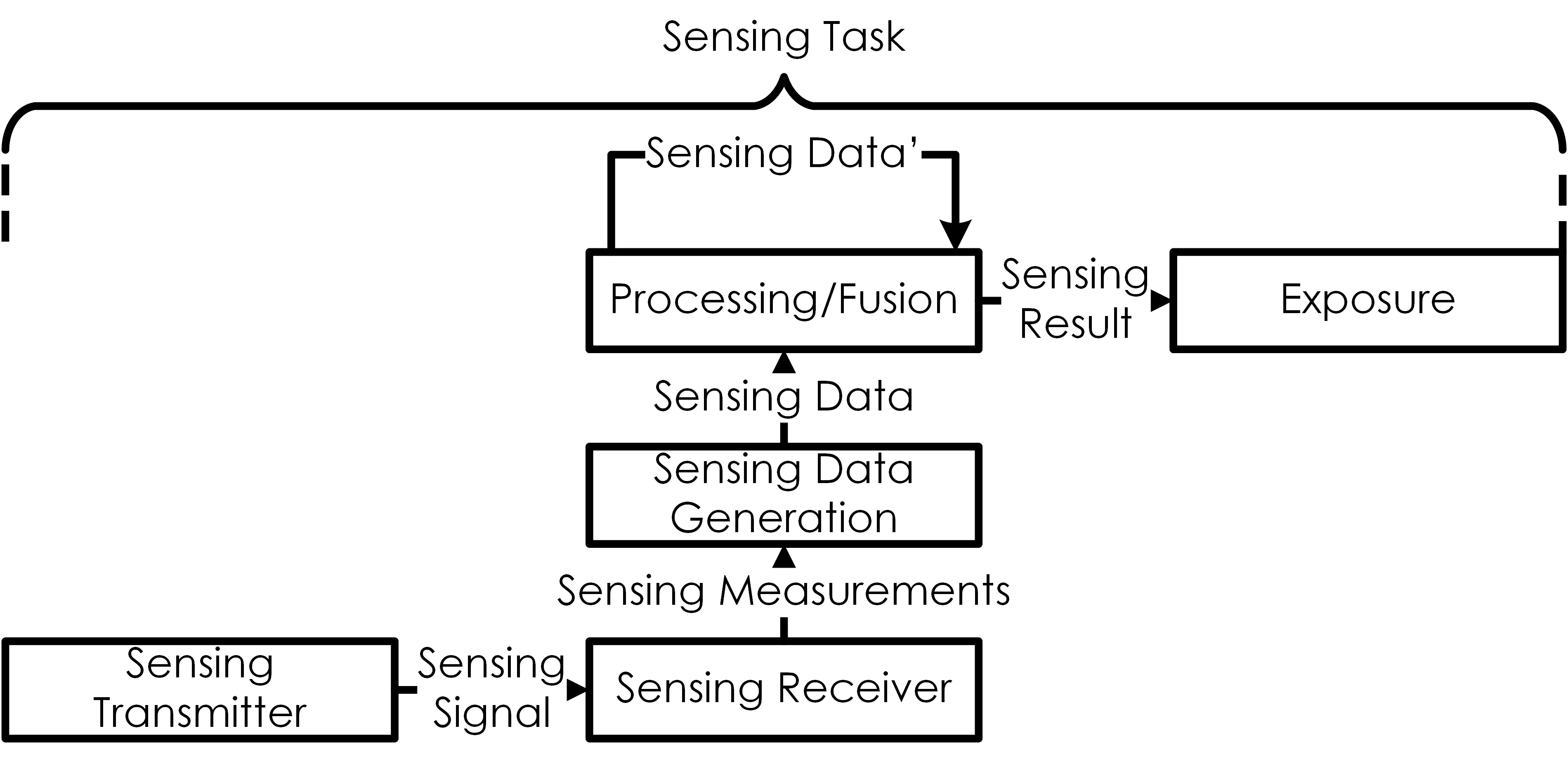}
\caption{Considered Sensing Workflow in \aclp{6GS}}
\label{fig:sensing-workflow}
\end{figure}

The workflow for sensing in a \ac{6GS} is depicted in \figurename~\ref{fig:sensing-workflow} and is based on \ac{ETSI}'s \ac{ISG} \ac{ISAC} \ac{GR} on use cases and requirements~\cite{etsi-isg-gr001}. The workflow shows functional blocks and the sensing information flow across them.

The functional blocks have the following meanings:
\\
\textbf{Sensing Signal} is a transmitted signal from a Sensing Transmitter for the purpose of sensing. The signal can be 6G or non-6G. Examples of a sensing signal are RF-based approaches or light for triggering sensors in cameras and \acsp{LiDAR}.
\\
A \textbf{\ac{STX}} is a 6G or non-6G entity that transmits a Sensing Signal. A \textbf{\ac{SRX}} is a 6G or non-6G entity that receives a Sensing Signal and produces sensing measurements. 
\\
An \textbf{\ac{SRX}} is a 6G or non-6G entity that receives a Sensing Signal. An \ac{SRX} can be co-located with an \ac{STX}.
\\
\textbf{Sensing data} is the 6G or non-6G data produced for sensing purposes.
\\
The \textbf{sensing data generation} step is a \ac{UE} or \acp{AN} functionality to take \ac{I/Q} or \ac{CSI} data from the \ac{SRX} and generate sensing data.
\\
A \textbf{sensing service} is a feature of the \ac{6GS} that is offered to service consumers. A Sensing Service provides sensing results based on communicated requirements and \acp{KPI}.
\\
A \textbf{\acl{ST}} consists of activities to perform sensing, including the configuration of the required \acp{STX} and \acp{SRX} (if applicable), the collection of sensing data, the processing of the sensing data and the exposure of the sensing results. Each \ac{ST} fulfils a Sensing Service request.
\\
A \textbf{\acl{TSSA}} is a 2- or 3-dimensional cartesian location area described with relative or absolute cartesian points that needs to be sensed by deriving characteristics of the environment and/or objects within the environment with certain sensing quality from the impacted (e.g. reflected, refracted, diffracted) Sensing Signals.
\\
The \textbf{sensing results} may include characteristics of objects (e.g. type, distance, velocity, trajectory, size, shape, material), or other contextual information (e.g. time of generation, environmental information) about objects in the \ac{TSSA}. Note, it is not precluded that the sensing result exposed to an entity within \ac{6GS} or to a third party may in some cases consist of the sensing data itself.
\\
\textbf{Fusion} refers to a process to join two or more streams of sensing data or sensing results together to form one or more sensing data or sensing result stream(s). Fusion can take place at the origin of the sensing data, or along the system entities of a \ac{6GS}, and in some cases in non-\ac{6GS} entities. The fusion of sensing results can also take place along all \ac{6GS} system entities (i.e. \ac{UE}, \ac{AN}, \ac{CN}).
\\
The \textbf{exposure} process coordinates the \ac{AAA} of third-party components to request and retrieve sensing results from the \ac{6GS}.

\subsection{Sensing-Enabled 6G \acl{SA}}\label{sec:archconsiderations-sa}
\begin{figure}[!t]
\centering
\includegraphics[width=0.85\linewidth]{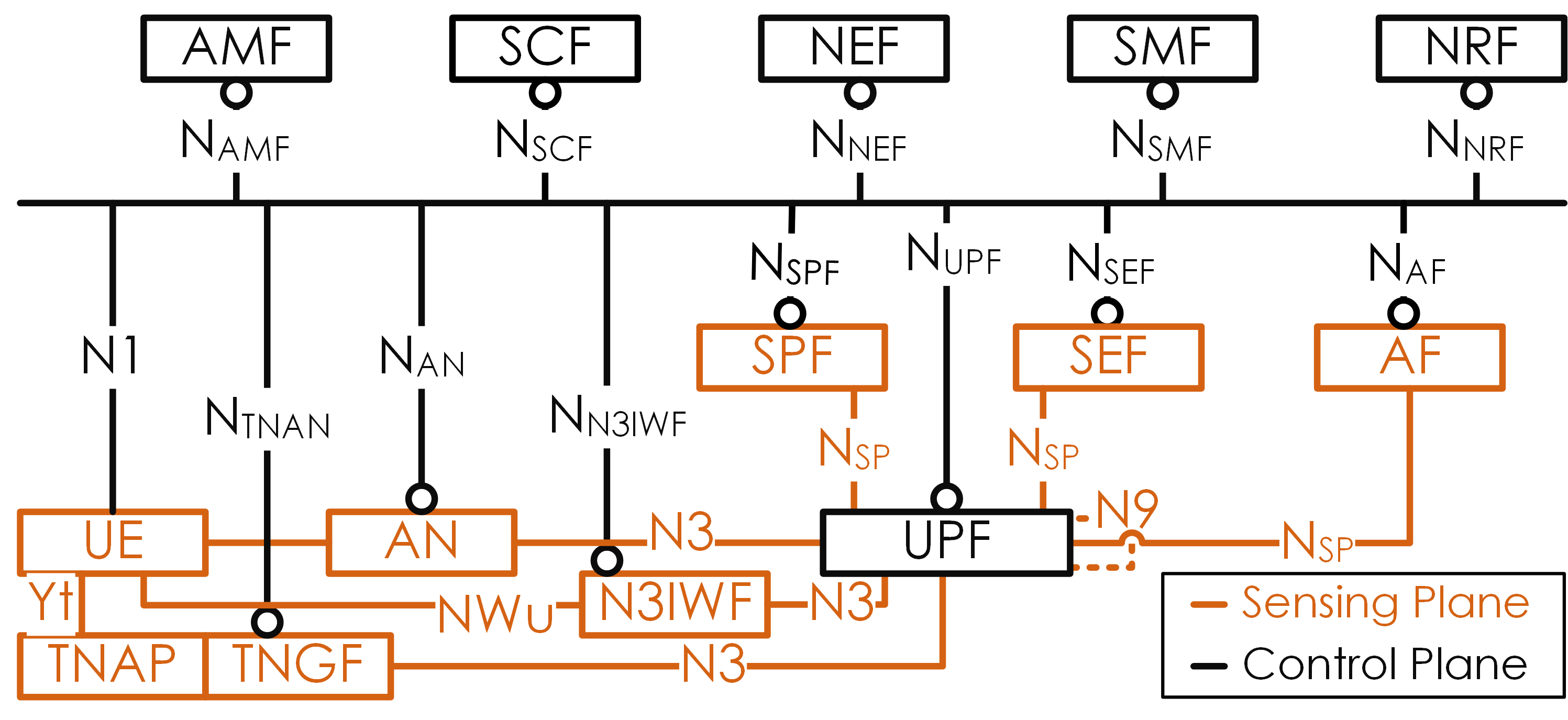}
\caption{Considered 6G System Architecture with support for 6G and non-6G Sensing.}
\label{fig:6g-arch}
\end{figure}

This section presents \acl{SA} considerations for an \ac{ISAC}-enabled \ac{6GS}. These considerations are rooted in on-going system research work of the \ac{EC}-funded MultiX project~\cite{multix}, with insights from the 3GPP (e.g. \cite{3gpp-6gws}), the recent 5G-A Stage 2 study~\cite{3gpp23700-14} and work within the \ac{ETSI} \ac{ISG} \ac{ISAC}.

\figurename~\ref{fig:6g-arch} illustrates \ac{SA} considerations on how to enable \ac{ISAC} in a \ac{6GS}. For readers well adversed in 5G, the illustrated  system architecture builds upon an \ac{SBA} proposition in the \ac{CN} where system functionality is disintegrated into different \acp{NF} for the purpose of multi-vendor deployments, the ability to scale on demand and the ability to choose not to deploy certain system functionality.

To enable \ac{ISAC}, there is no difference to this proposition and new \acp{NF} are proposed for 6G. Before delving into their purpose, one may notice the disappearance of the \ac{AMF} from the \ac{SA}; with a range of key contributions to 3GPP calling out the limitations of deploying the \ac{CN} in a truly cloud-native fashion where no \ac{NF} has a mix of service- and non-service-based interfaces, e.g. the \ac{AMF}. As a result, the \ac{SBA} bus is extended down to the \ac{AN}, enabling an \ac{SBI} towards the \ac{AN}, i.e. $N_{AN}$, and the reduction of the \ac{AMF}'s functionality to exclude the proxy-behaviour of all \ac{BS}-\ac{CN} traffic. Furthermore, based on recent advances in 3GPP on upgrading the \ac{UPF} to expose monitoring information to other \ac{NF} via an \ac{SBI}~\cite{3gpp-upeas}, \figurename~\ref{fig:6g-arch} adopts this view, leaving options open for the \ac{UPF} to be configured via $N4$, but exposes monitoring information via $Nupf$.

The newly proposed \ac{ISAC}-related \acp{NF} are listed below in alphabetical order and described in their functionalities for the purpose of offering a \ac{6GS} sensing service and realising the sensing workflow described in Section~\ref{sec:archconsiderations-sensingworkflow}:

\textbf{\ac{SCF}:} This \ac{NF} is responsible for the coordination of establishing, modifying and terminating \acp{ST}. \ac{SCF} is coordinating this in close relationship with the \ac{AN} to select the \ac{STG} members (\acp{UE}, \acp{AN}, \acp{AF}) based on their capabilities, availability and utilisations for 6G and non-6G sensing sources. This includes assistant information towards the \ac{AN} on selecting the most appropriate sensing mode for the successful \ac{ST} execution. \ac{SCF} also manages sensing mobility of target objects while working closely with the \ac{AMF} for \ac{UE} mobility scenarios where the \ac{UE} is either an \ac{STX}, \ac{SRX} or both.

\textbf{\ac{SPF}:} This \ac{NF} collects, processes, fuses and stores sensing data from \acp{SRX}. The processing and fusion may leverage \ac{AI} techniques and the \ac{SPF} may offer capabilities for model training, storage and inference. The \ac{SPF} outputs sensing results.
    
\textbf{\ac{SEF}:} This \ac{NF} is responsible for exposing sensing results from the \ac{SPF} to \ac{6GS} third party components. Key \ac{AAA} functionality also resides in the \ac{SEF} for \ac{ISAC}-related purposes.

Non-6G \acp{AN} are of great use to complement limitations of sensing-enabled 6G \acp{AN}, as described in the use case provided in Section~\ref{sec:usecases-uc1}. To enable the \ac{6GS} to coordinate Wi-Fi sensing of trusted and untrusted non-6G \acp{AN}, \figurename~\ref{fig:6g-arch} illustrates the 5G components \ac{TNAN} and \ac{N3IWF} offering an \ac{SBI} of their own for the \ac{SCF} to leverage them to execute an \ac{ST}.

Besides extending the \ac{SBA} to all system components, apart from the \ac{UE},  \figurename~\ref{fig:6g-arch} also illustrates a \ac{SP}. In comparison to the traditional \ac{UP} between the \ac{UE} and the \ac{DN} via the \ac{AN} and \ac{UPF}, the \ac{SP} allows sensing data, i.e. processed \ac{I/Q} data from \ac{SRX} - to be exchanged between system components that either produce sensing results (i.e. the \ac{SPF}) and share/expose them (i.e. \ac{SPF} and optionally \ac{SEF} depending on the \ac{SSC}) or forward the sensing data to system components (i.e. \ac{SPF}) to process and fuse sensing data for the purpose of generating sensing results eventually. As the system components to achieve the generation of sensing results are not in the \ac{DN}, the need for a dedicated \ac{SP} is deemed necessary to enable \ac{QoS} enforcement over the exchange of sensing data and sensing results to meet the requested \acp{KPI} for the sensing service. The proposed \ac{SP} may utilise the 5G tunnelling protocol \acs{GTP-U} or any other stateless protocol such as \ac{SRv6} \cite{rfc8986} or \ac{MASQUE}~\cite{ietf-masque}. It is proposed to utilise the stateless \acs{HTTP/3} protocol, allowing all endpoints in the \ac{SP} to leverage \ac{SBA} principles.

\begin{figure}[!t]
\centering
\includegraphics[width=0.75\linewidth]{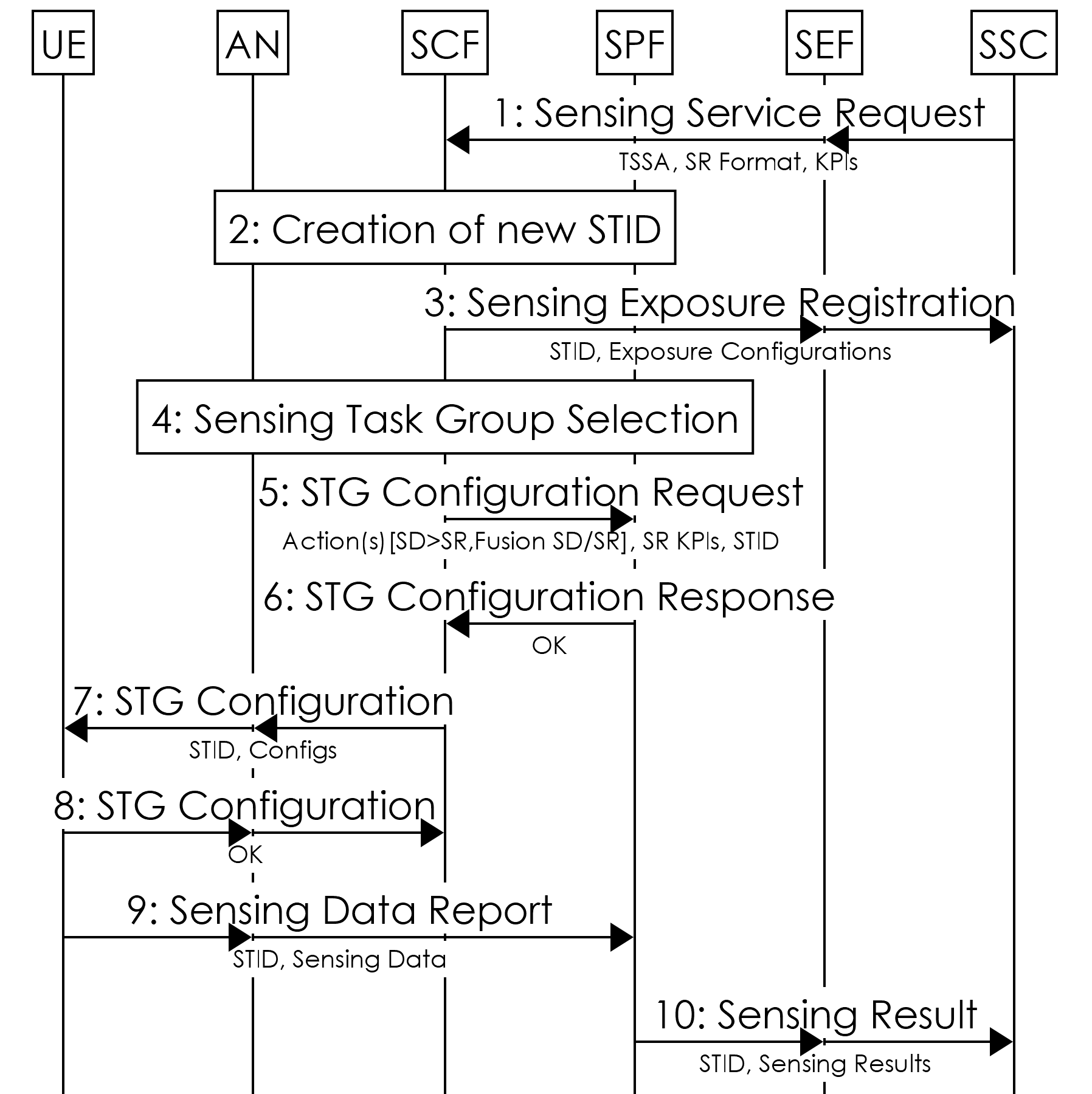}
\caption{End-to-end call flow for \ac{ST} execution in \aclp{6GS}}
\label{fig:6g-st-execution}
\end{figure}

To further illustrate how an \ac{ST} is executed by the \ac{6GS}, \figurename~\ref{fig:6g-st-execution} depicts an end-to-end call flow where an \ac{SSC} (6G internal system or third-party components) requests a sensing service from the \ac{6GS} which goes to the \ac{SCF}, optionally via the \ac{SEF} if the \ac{SSC} is a third-party where \ac{AAA} must be ensured. \ac{SCF} then creates a random but system-wide unique \ac{STID} and sends it back to the \ac{SSC} (via the \ac{SEF} if requested by a third party). The \ac{SCF} then determines the \ac{STG} members (i.e. \ac{STX}(s), \ac{SRX}(s) and \ac{SPF}(s)) for the purpose of sensing in a \ac{TSSA} and prepares all selected \ac{SPF} instances with the necessary configurations - including the requested sensing result \acp{KPI}) to process and/or fuse sensing data from \acp{SRX}; the generated \ac{STID} is also shared with the \ac{SPF}. To kick-off the sensing activity, the \ac{SCF} configures all \ac{STG} members that have been identified as \ac{STX} and/or \ac{SRX} in Step 4. Once \ac{ST} is active, the \ac{SRX} reports the generated sensing data to an \ac{SPF} for processing/fusion and the \ac{SRX} includes the \ac{STID}, allowing the \ac{SPF} to check what is required from it using the configurations from the \ac{SCF} in Step 5. The sensing results are then sent/exposed to the \ac{SSC}. 

\subsection{\acl{CP} Considerations for ISAC in 6G}
\begin{figure*}[!t]
\centering
\includegraphics[width=0.89\linewidth]{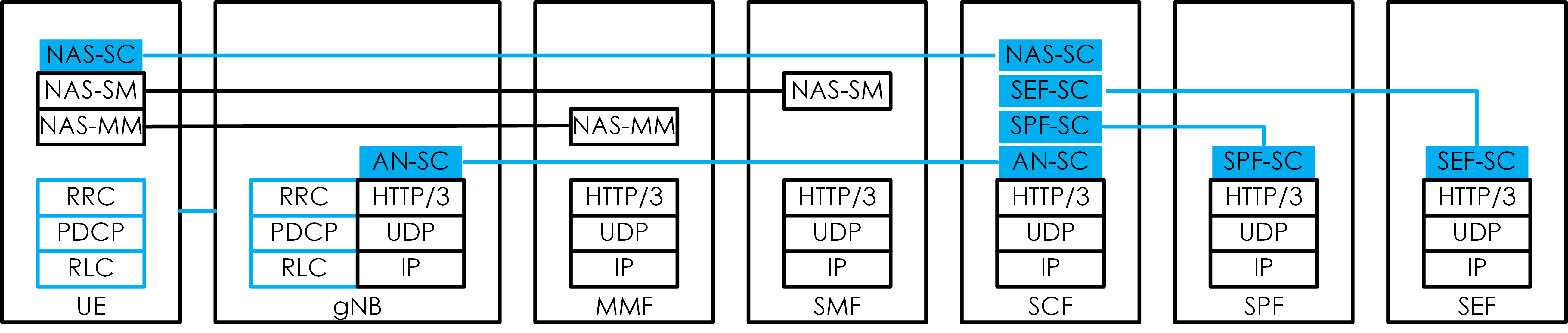}
\caption{Considered Sensing-Specific \acl{CP} for \aclp{6GS}.}
\label{fig:6g-cp}
\end{figure*}

The considered \ac{CP} for \ac{ISAC}-enabled \ac{6GS} is illustrated in \figurename~\ref{fig:6g-cp} and depicts the protocol stack for the considered system architecture in Section~\ref{sec:archconsiderations-sa}. As indicated in that section, the \ac{UE} continues using the 5G $N1$ procedures (i.e. \ac{NAS}) but over an \ac{SBA} service bus. This allows the simplification of the protocol stack whereby \acp{UE} can communicate with any \ac{NF} directly; not only do \acp{UE} have direct \ac{MM} and \ac{SM} signalling to the \ac{AMF} and \ac{SMF}, the same applies for any \ac{SC} \ac{CP} signalling between the \ac{UE} and the \ac{SCF}. The \ac{SC} signalling allow the \ac{UE} to register its sensing capabilities with the \ac{CN} and the \ac{SCF} to configure the \ac{UE} to execute any of the sensing workflow functionalities described in Section~\ref{sec:archconsiderations-sensingworkflow}. 

In the \ac{RAN}, various sensing functionality is enabled across the protocol stack and it is on-going research in standards in general and the MultiX project in particular to design which layer implements which \ac{SC} functionality. Hence, all layers are illustrated to receive potential changes (shown in light-blue).

With the \ac{SA} change to upgrade the 5G $N2$ interface (\ac{NGAP} over \ac{SCTP}) to an \ac{SBI}, \acp{AN} (i.e. \ac{gNB}, \ac{TNAN} and \ac{N3IWF}) can also directly exchange \ac{CP} signalling with the \ac{SCF} and do so over \ac{HTTP/3} which leverages \ac{UDP}, similar to how \ac{SCTP} operates to avoid any blocking behaviour such as \ac{TCP} would cause in congestion scenarios. To further allow for modularity and enforce unification and cloud-nativeness of any \ac{NF} in the \ac{6GS}. This results in the \acp{SBI} \acs{AN-SC} for the \ac{AN}. As the \ac{ST} spans over the processing and fusion of sensing data, the \ac{SCF} has an \ac{SBI} towards the \ac{SPF}, denoted as $\acs{SPF-SC}$ in \figurename~\ref{fig:6g-cp} and $Nspf$ in \figurename~\ref{fig:6g-arch}. For exposure purposes, the same logic applies in manifestation of the $\acs{SEF-SC}$ interface between the \ac{SCF} and the \ac{SEF}.


\section{Conclusions}\label{sec:conclusions}
As 6G continues to evolve, \ac{ISAC} is poised to become a cornerstone of next-generation mobile systems, facilitating not only communication but also environmental awareness across diverse application domains. This paper presented key architectural and protocol enhancements required to support \ac{ISAC} in 6G, grounded in real-world use cases from the MultiX project. By identifying functional requirements and proposing dedicated network functions and planes, a cohesive vision of how ISAC can be systematically integrated into a cloud-native, service-based 6G \ac{CN} is provided. The proposed architecture emphasises modularity to support advanced sensing services while maintaining compatibility with both 6G and non-6G sensing technologies. These insights aim to guide future standardisation efforts and support the scalable deployment of \ac{ISAC}-enabled services in 6G systems.

\bibliographystyle{IEEEtran}
\bibliography{IEEEabrv,references.bib}

\end{document}